\documentclass[conference]{IEEEtran}
\IEEEoverridecommandlockouts
\pdfoutput=1
\usepackage{cite}
\usepackage{amsmath,amssymb,amsfonts}
\usepackage{algorithm}
\usepackage{algpseudocode}
\usepackage{graphicx}
\usepackage{textcomp}
\usepackage{ifthen}
\usepackage{bbm}
\usepackage{xcolor}
\usepackage{tikz}
\usepackage{graphicx}
\usepackage{subcaption}
\usepackage[square,numbers]{natbib}
\usepackage{url}            
\usetikzlibrary{automata, positioning}
\def\BibTeX{{\rm B\kern-.05em{\sc i\kern-.025em b}\kern-.08em
    T\kern-.1667em\lower.7ex\hbox{E}\kern-.125emX}}
\newcommand{\Exp}[1]{\mathbb{E}\left[#1\right]}

\newcommand{\ra}{\rightarrow}

\newcommand{\floor}[1]{\left\lfloor #1 \right\rfloor}

\newcommand{\N}{\mathbb{N}} 
 
\newcommand{\ignore}[1]{}

\newboolean{showcomments}
\setboolean{showcomments}{true}
\newcommand{\ak}[1]{  \ifthenelse{\boolean{showcomments}}
{ \textcolor{red}{(AK says:  #1)}} {}  }
\newcommand{\jk}[1]{  \ifthenelse{\boolean{showcomments}}
{ \textcolor{red}{(JK says:  #1)}} {}  }

\newtheorem{theorem}{Theorem}

\newtheorem{lemma}{Lemma}

\allowdisplaybreaks[4]
\begin{document}

\title{``Please come back later'': Benefiting from deferrals in
  service systems}

\author{\IEEEauthorblockN{Anmol Kagrecha}
\IEEEauthorblockA{\textit{EE, IIT Bombay} \\
\textit{akagrecha@ee.iitb.ac.in}}
\and
\IEEEauthorblockN{Jayakrishnan Nair}
\IEEEauthorblockA{\textit{EE, IIT Bombay} \\
\textit{jayakrishnan.nair@ee.iitb.ac.in}}}
\maketitle

\begin{abstract}
The performance evaluation of loss service systems, where customers
who cannot be served upon arrival get dropped, has a long history
going back to the classical Erlang~B model. In this paper, we consider
the performance benefits arising from the possibility of deferring
customers who cannot be served upon arrival. Specifically, we consider
an Erlang~B type loss system where the system operator can, subject to
certain constraints, ask a customer arriving when all servers are
busy, to \emph{come back} at a specified time in the future. If the
system is still fully loaded when the deferred customer returns, she
gets dropped for good. For such a system, we ask: How should the
system operator determine the `rearrival' times of the deferred
customers based on the state of the system (which includes those
customers already deferred and yet to arrive)? How does one quantify
the performance benefit of such a deferral policy?

Our contributions are as follows. We propose a simple state-dependent
policy for determining the rearrival times of deferred customers. For
this policy, we characterize the long run fraction of customers
dropped. We also analyse a relaxation where the deferral times are
bounded in expectation. Via extensive numerical evaluations, we
demonstrate the superiority of the proposed state-dependent policies
over naive state-independent deferral policies.

\end{abstract}


\section{Introduction}
\label{sec:intro}

Many service systems have the property that denial of service upon the
arrival of a request results in the request getting dropped.  The
classical example comes from telephony, where an incoming call request
must either be connected, or dropped. Another contemporary example is
an electric car charging facility, which cannot serve an incoming
customer when all charging stations are occupied. Similarly, certain
online services decline fresh requests when congested.

Such service systems, by their very nature, are unable to \emph{queue}
requests that arrive when service capacity is fully utilized. It is
therefore natural to ask: What if these systems were instead able to
\emph{defer} some requests, i.e., have these requests `come back' at a
pre-specified time in the future? If so, what is optimal strategy for
deferring requests? To what extent does the throughput of the system
improve from such a deferral policy? The goal of this paper is to shed
light on these questions.

Formally, we consider an Erlang B (M/M/$K$/$K$) service system, where
the system operator can, subject to certain constraints, defer jobs
that arrive when all servers are occupied. The constraints include an
upper bound on the deferral time (which is natural given QoS
considerations), and/or a constraint on the number of jobs that can be
deferred at any time. In this setting, our goal is to design simple
deferral policies that are effective, while also being analytically
tractable. The main difficulty in the performance evaluation of
deferral policies is that the state description must incorporate
information about (future) arrival instants of deferred jobs.

Our main contribution is the design and analysis of a deferral policy,
which spaces arrivals of deferred jobs uniformly in the future. Even
though this policy induces a complicated (uncountable) Markovian state
space, we are able to analytically characterize the blocking
probability via a novel combination of steady state and transient
analysis of certain Markov processes. Via numerical experiments, we
show that the proposed policy outperforms naive state-independent
deferral policies, as well as the Erlang B system (which does not
allow deferrals).

A relaxation on the deferral time constraint is considered next, where
the upper bound on the deferral time is imposed \emph{in
expectation}. For this relaxed model, we propose a deferral policy,
which admits a considerably more explicit performance
characterization. Via numerical experiments, we show that the
performance of this policy is in fact a good approximation to the
policy designed for the `hard' deferral constraint.

The remainder of the paper is as follows. We provide a brief survey of
related literature next. In Section~\ref{sec:model}, we describe the
primary system model and state some preliminary results. The `hard'
case where deferral times are bounded deterministically is considered
in Section~\ref{sec:sd_policy}, and the relaxed model is considered in
Section~\ref{sec:model2}.

\textit{Related literature}: This work is motivated by the recent work
\cite{doshi2015}, which implemented a deferral system for a web-server
that is prone to congestion. To the best of our knowledge, there is no
prior work that provides an analytical treatment of queueing systems
that allow deferral.

This work is peripherally related to the considerable literature on
retrial queues (we refer the reader to a recent survey paper
\cite{kim2016} and the book \cite{artalejo1999}). Retrial queues are
systems where blocked customers `try again later'. The main contrast
between retrial queues and deferral queues is that in the former
setting, the retrial model is taken to be exogenous and indicative of
customer behavior, while in the latter setting, deferral is a control
decision on part of the system operator.

Another related class of loss systems that have been studied are
abandonment based queues (references include \cite{baccelli1981} and
\cite{zeltyn2005}). In these systems, customers who are denied service
upon arrival wait for a certain (potentially random) amount of
time. If service is not offered before this patience time runs out,
the customer leaves. In contrast, in deferral queues, a deferred
customer does not actually wait, but rather leaves the system to come
back at a pre-specified time. So this deferred customer can only be
served if service capacity is available at the precise moment of
re-entry. In this sense, abandonment based systems provide lower
bounds on the performance achievable in deferral systems.

\section{Model and Preliminaries}
\label{sec:model}

We consider a loss service system consisting of $K$ identical parallel
servers of unit speed. Jobs (or customers) arrive according to a
Poisson process of rate $\lambda,$ and service times are exponentially
distributed with rate $\mu.$ In the absence of deferrals, this system
is the classical Erlang~B (M/M/$K$/$K$) queue.

On this baseline Erlang~B model, we allow the system operator (or the
scheduling policy) to \emph{defer} jobs that arrive when all servers
are busy, to come back to the system at a pre-specified time in the
future. If a free server is available when this deferred job comes
back, it gets admitted into service. If not, the job gets dropped. In
other words, we only allow a job to be deferred once. We refer to the
time instant at which a deferred job comes back to the system as its
\emph{rearrival time}, and we refer to the interval between the
rearrival time and the original arrival time as the \emph{deferral
time}.

Of course, it is natural to impose certain constraints on the deferral
policy. Firstly, QoS considerations dictate that deferral time cannot
be too large. We impose this constraint in two ways: In
Section~\ref{sec:sd_policy}, we consider the case where there is a
deterministic upper bound $\hat{T} > 0$ on deferral times. In
Section~\ref{sec:model2}, we consider a relaxed model where the
deferral times are bounded above \emph{in expectation} by $\hat{T}.$ A
second constraint we impose is an upper bound $\hat{D}$ on the number of
deferred jobs at any time. Finally, we impose \emph{work
conservation}, i.e., when a job arrives (either for the first time, or
post deferral), it must be served so long as there is at least one
free server available.

Our metric for evaluating the performance of a deferral policy is the
long run fraction of jobs blocked, a.k.a., the blocking
probability. Note that jobs can be blocked in two ways: One, a job
arrives when all servers are busy, and there are already $\hat{D}$
previous arrivals that have been scheduled for rearrival in the
future. Second, a deferred job on rearrival finds that all the servers
are (still) busy.

A naive deferral policy would be to assign a large deferral time (say
$\hat{T}$), subject to limit $\hat{D}$ on number of deferred jobs.
Another naive policy would be to assign a deferral time sampled
uniformly at random in $[0,\hat{T}],$ again subject to the deferral
limit. Such policies are \emph{state-independent}, in that the
deferral time of a job is assigned independently of the deferral times
assigned to previous jobs.
In this paper, we propose low overhead \emph{state-dependent} deferral
policies, that schedule rearrivals of deferred jobs cognizant of the
already scheduled rearrivals.

Finally, we note that work conservation ensures that the blocking
probability of any deferral policy is at most the blocking probability
in the Erlang~B system. In other words, the Erlang~B formula provides
an upper bound on the blocking probability of any deferral policy.

Throughout the paper, for $n \in \N,$ the set $\{0,1,2,\cdots,n\}$ is
denoted by $[n].$

\section{Bounded Deferral Times}
\label{sec:sd_policy}

In this section, we consider the case where the deferral times are
bounded by $\hat{T} > 0,$ i.e., the system cannot defer an incoming job by
more than $\hat{T}$ time units.\footnote{While we take $\hat{T}$ to be an
  exogenous model parameter, its value may in practice be determined
  based on QoS considerations and customer patience.} For this model,
we propose a simple state-dependent deferral policy, which we refer to
as the {\bf D}eterminstically {\bf S}paced {\bf R}earrival {\bf T}imes
(DSRT) policy, which spreads the rearrival instants of deferred jobs
`uniformly', so as to maximize the chances of admitting them. We
provide a characterization of the blocking probability under this
policy, via a novel combination of the steady state analysis of a
Markov chain that captures the system evolution in the absence of
deferrals, and the transient analysis of another Markov chain that
captures the system evolution in the presence of deferrals. Finally,
we present a case study that demonstrates the superiority of our
state-dependent deferral policy over naive state-independent policies.

\subsection{DSRT Policy}

The proposed deferral policy is described as follows. It is
parameterized by $x \in (0,\hat{T}],$ which specifies the spacing
between arrivals of deferred jobs into the system.
If there are no jobs currently deferred, and an arriving job finds all
servers busy, then the policy defers that job by $x$ time units. On
the other hand, if there are deferred jobs, and an arriving job finds
all servers busy, then the policy schedules its rearrival time to be
$x$ time units after the rearrival time of the last deferred job,
subject to a maximum of $D=\min(\hat{D}, \floor{\hat{T}/x})$ deferred
jobs at any time.
This policy is defined formally as Algorithm~\ref{alg:sddp}.


\begin{algorithm}
  \caption{DSRT policy}
  \label{alg:sddp}
\begin{algorithmic}
  \Procedure{DSRT}{}
  \If{$\text{num\_deferrals} = 0$}
    \State $\text{rearrival\_time} \gets \text{arrival\_time} + x$
  \ElsIf{$\text{num\_deferrals} < D:=\min(\hat{D}, \floor{\hat{T}/x})$}
    \State $\text{rearrival\_time} \gets \text{prev\_rearrival\_time} + x$
  \Else
    \State $\text{Customer blocked}$
  \EndIf
  \EndProcedure
\end{algorithmic}
\end{algorithm}

Note that the proposed DSRT policy attempts to spread the rearrival
instants $x$ units of time apart, subject to the deferral constraints.
Intuitively, if the spacing $x$ is too small, then the deferred jobs
are likely to find all servers occupied upon rearrival. On the other
hand, if $x$ is too large, the policy is inefficient, since fewer jobs
may be deferred at any time (given the constraint $\hat{T}$ on the deferral
times), and fewer opportunities are created to admit deferred
jobs. Thus, the policy must operate at an intermediate `soft spot'; we
address this issue in our numerical experiments.

Under the DSRT policy, the temporal evolution of the system can be
captured as a Markov process over state space
$[K] \times [D] \times [0,x].$ Here, the state $(k,d,\theta)$
indicates that $k$ servers are currently busy, there are $d$ jobs in
deferral, and that the earliest rearrival of a deferred job will occur
after time $\theta.$ Note that the state space is uncountable, making
an analysis of the stationary distribution cumbersome. We overcome
this difficulty by directly characterizing $\Pi_{k,d}$ which is the
long run fraction of time the state is of the form $(k,d,\theta)$ for
some $\theta \in [0,x].$ This is done by separately analysing the
temporal evolution of the system in the presence and absence of
deferred jobs, and then combining these analyses in a novel manner.

We begin by considering the special case of a single server, and a
limit of at most one deferral at any time ($K = D = 1$). This special
case is instructive, not just because it illustrates our analysis
approach, but also because it admits a closed form characterization of
the blocking probability. We then consider the case of general $K,$
$D$ in Section~\ref{sec:mutiservermultideferral}; in this case, our
characterization is less explicit, though amenable to an exact
computation.

\subsection{Single-Server, Single-Deferral System}

\begin{figure}[htp]
    \centering
    \includegraphics[width=\linewidth]{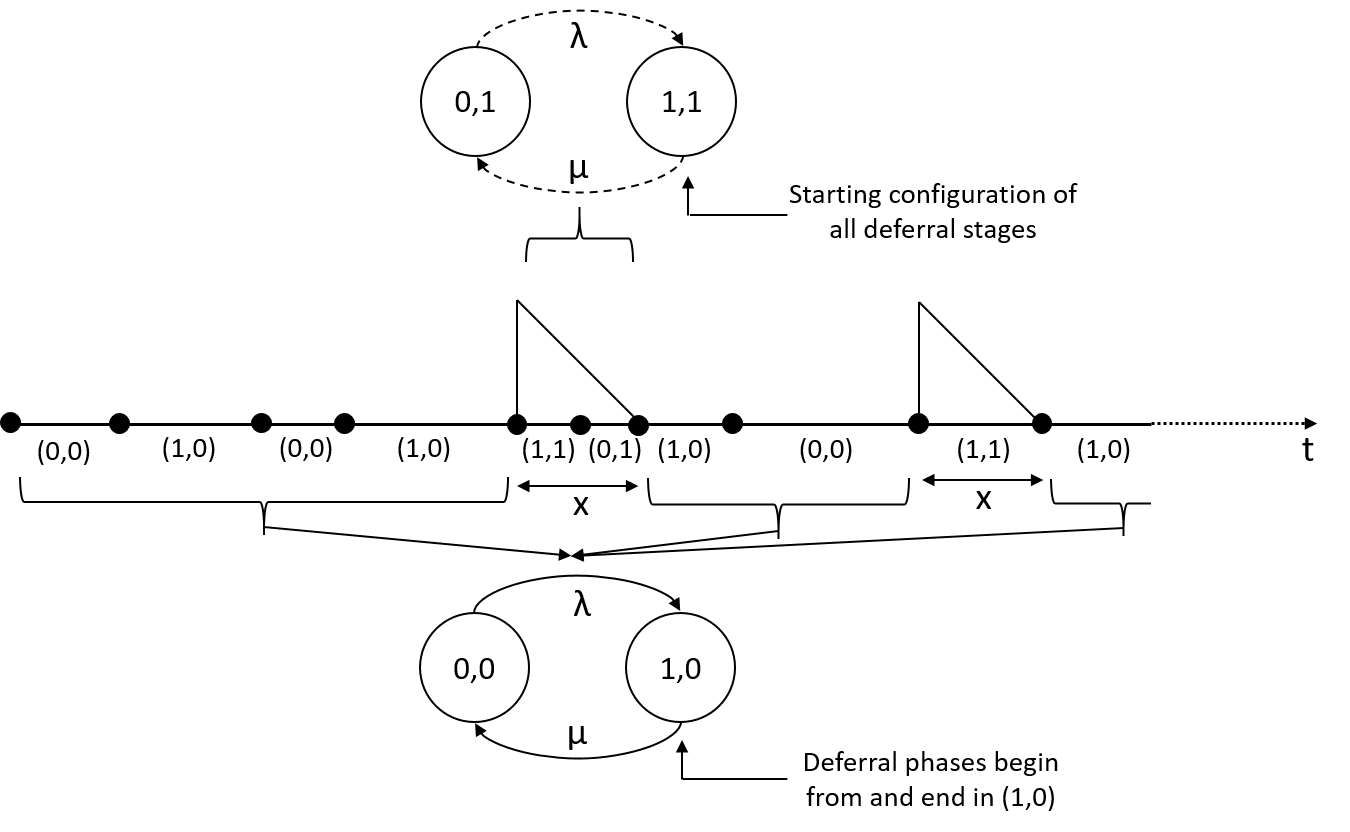}
    \caption{Illustration of the temporal evolution of $\theta$ when
      $K = D = 1.$ Deferral phases ($d > 0$) of length $x$ are
      interspaced between non-deferral phases ($d = 0$).}
    \label{fig:dyanmics_fig}
\end{figure}  

The temporal evolution of the system for the case $K = D = 1$ is
illustrated in Figure~\ref{fig:dyanmics_fig}. The key observation that
informs our analysis is the following. Considering only the times when
the system has no deferrals (i.e., disregarding the deferral phases,
when $d >0$), the system configuration evolves as per the Markov chain
depicted at the bottom of Figure~\ref{fig:dyanmics_fig}.\footnote{We
  avoid referring to $(k,d)$ as the \emph{state} of the system, and
  instead refer to it as the \emph{configuration}.} On the other hand,
over each \emph{deferral phase} of length $x$ (when $d > 1$) the
system configuration evolves as per the Markov chain depicted at the
top of Figure~\ref{fig:dyanmics_fig}.

Since the system configuration evolves as per the Markov chain
depicted at the bottom of Figure~\ref{fig:dyanmics_fig} when there are
no deferrals, the ratio between $\Pi_{0,0}$ and $\Pi_{1,0}$ is
dictated by the flow balance equations for that Markov chain:
\begin{equation}
	\label{eq:ss-flow-1}
	\lambda \Pi_{0,0} = \mu \Pi_{1,0}.
\end{equation}
      
Turning now to the configurations where $d = 1,$ we note that the
relationship between $\Pi_{0,1}$ and $\Pi_{1,1}$ is dictated by the
transient behavior of the Markov chain depicted on the top of
Figure~\ref{fig:dyanmics_fig} over an interval of length $x,$ starting
in configuration $(1,1).$ Focusing on this Markov chain over the
interval $[0,x],$ the distribution of the chain at time $t,$ captured
by $\textbf{p}(t) = [p_{0,1}(t), p_{1,1}(t)],$ is given by:
\begin{equation*}
 	\textbf{p}(t) = [0,1] e^{Qt},
\end{equation*}
where 
\begin{equation*}
	Q = 
	\begin{bmatrix}
		-\lambda & \lambda \\
		\mu & -\mu
	\end{bmatrix}.
\end{equation*}
It now follows that       
\begin{equation}
	\label{eq:ss-fraction}
	\frac{\Pi_{0,1}}{\Pi_{1,1}}
	= \frac{\int_{0}^{x} p_{0,1}(t) dt}{\int_{0}^{x} p_{1,1}(t) dt}.
\end{equation}

Next, we need to relate the long run fraction of time spent in
non-deferral configurations $(d = 0)$ to those under deferral
configurations $(d = 1)$. To do this, we note, using the PASTA
property, that the rate at which transitions occur from
non-deferral configurations to deferral configurations equals $\lambda
\Pi_{1,0}.$ On the other hand, transitions from deferral
configurations to non-deferral configurations occur at rate
$\frac{\Pi_{0,1} + \Pi_{1,1}}{x}.$ Equating the two, we get:
\begin{equation}
	\label{eq:ss-flow-2}
 	\lambda \Pi_{1,0} = \frac{\Pi_{0,1} + \Pi_{1,1}}{x}. 
\end{equation}
      
Lastly, we have the normalization condition
\begin{equation}
	\label{eq:ss-tpr}
	\Pi_{0,0} + \Pi_{1,0} + \Pi_{0,1} + \Pi_{1,1} = 1.
\end{equation}
Solving the system of equations
\eqref{eq:ss-flow-1}--\eqref{eq:ss-tpr} gives yields a closed form
characterization of the long run fraction of time spent in each
configuration. 

\begin{figure*}
  \centering
  \includegraphics[width=0.8\linewidth]{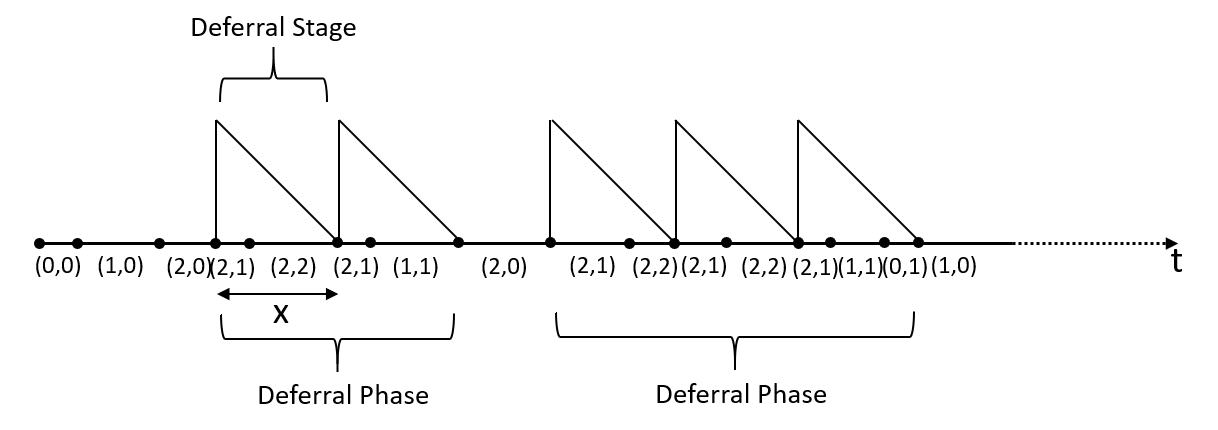}
  \caption{Illustration of the temporal evolution of $\theta$ when
    $K = 2, D = 3.$}
  \label{fig:dyanmics_fig_2}
\end{figure*}

\begin{figure}
\centering
\begin{tikzpicture}[state/.style={circle, draw, minimum size=1.4cm}]
\node[state] (zero-zero) {0,0};
\node[state, right=of zero-zero] (one-zero) {1,0};
\node[state, right=of one-zero] (two-zero) {2,0};
\node[state, right=of two-zero] (k-zero) {K,0};

\path (two-zero) -- node[auto=false]{\ldots} (k-zero);

\draw[every loop]
  (zero-zero) edge[bend right, auto=left] node{$\lambda$} (one-zero)
  (one-zero) edge[bend right, auto=left] node{$\mu$} (zero-zero)
  (one-zero) edge[bend right, auto=left] node{$\lambda$} (two-zero)
  (two-zero) edge[bend right, auto=left] node{$2\mu$} (one-zero)
  
  (k-zero) edge[bend right, auto=left] node{$\lambda \alpha_{2}$} (two-zero)
  (k-zero) edge[bend right, auto=right] node{$\lambda \alpha_{1}$} (one-zero);
\end{tikzpicture}
\caption{System evolution over the non-deferral configurations}
\label{fig:msmd_ctmc}
\end{figure}
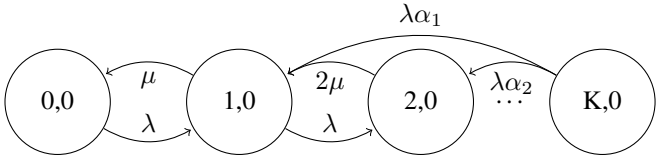

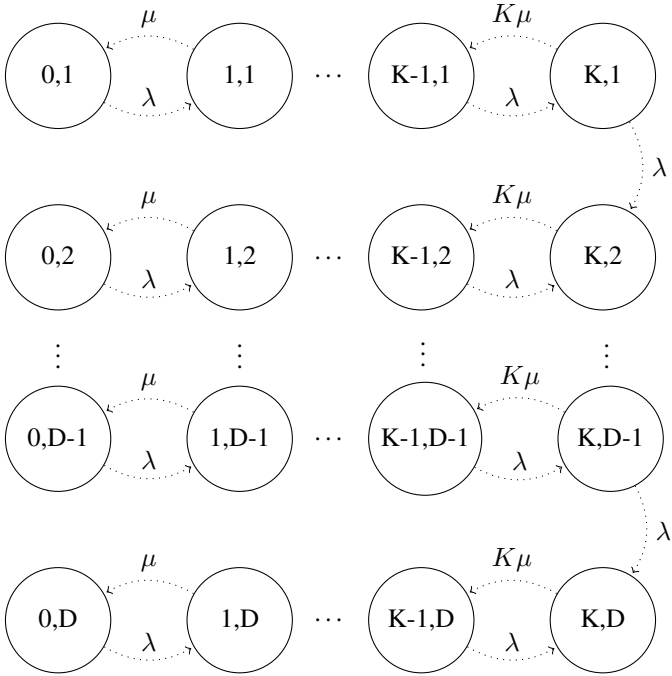
\begin{figure}
\centering
\begin{tikzpicture}[state/.style={circle, draw, minimum size=1.4cm}]
  \node[state] (zero-one) {0,1};
  \node[state, right=of zero-one] (one-one) {1,1};
  \node[state, right=of one-one] (km1-one) {K-1,1};
  \node[state, right=of km1-one] (k-one) {K,1};

  \node[state, below=of zero-one] (zero-two) {0,2};
  \node[state, below=of one-one] (one-two) {1,2};
  \node[state, right=of one-two] (km1-two) {K-1,2};
  \node[state, right=of km1-two] (k-two) {K,2};

  \node[state, below=of zero-two] (zero-dm1) {0,D-1};
  \node[state, below=of one-two] (one-dm1) {1,D-1};
  \node[state, right=of one-dm1] (km1-dm1) {K-1,D-1};
  \node[state, right=of km1-dm1] (k-dm1) {K,D-1};

  \node[state, below=of zero-dm1] (zero-d) {0,D};
  \node[state, below=of one-dm1] (one-d) {1,D};
  \node[state, right=of one-d] (km1-d) {K-1,D};
  \node[state, right=of km1-d] (k-d) {K,D};

  \path (one-one) -- node[auto=false]{\ldots} (km1-one);
  \path (one-two) -- node[auto=false]{\ldots} (km1-two);
  \path (one-dm1) -- node[auto=false]{\ldots} (km1-dm1);
  \path (one-d) -- node[auto=false]{\ldots} (km1-d);

  \path (zero-two) -- node[auto=false]{\vdots} (zero-dm1);
  \path (one-two) -- node[auto=false]{\vdots} (one-dm1);
  \path (km1-two) -- node[auto=false]{\vdots} (km1-dm1);
  \path (k-two) -- node[auto=false]{\vdots} (k-dm1);

  \draw[every loop]
    (zero-one) edge[bend right, auto=left, dotted] node{$\lambda$} (one-one)
    (one-one) edge[bend right, auto=right, dotted] node{$\mu$} (zero-one)
    (km1-one) edge[bend right, auto=left, dotted] node{$\lambda$} (k-one)
    (k-one) edge[bend right, auto=right, dotted] node{$K \mu$} (km1-one)
    (k-one) edge[bend left, auto=left, dotted] node{$\lambda$} (k-two)

    (zero-two) edge[bend right, auto=left, dotted] node{$\lambda$} (one-two)
    (one-two) edge[bend right, auto=right, dotted] node{$\mu$} (zero-two)
    (km1-two) edge[bend right, auto=left, dotted] node{$\lambda$} (k-two)
    (k-two) edge[bend right, auto=right, dotted] node{$K \mu$} (km1-two)

    (zero-dm1) edge[bend right, auto=left, dotted] node{$\lambda$} (one-dm1)
    (one-dm1) edge[bend right, auto=right, dotted] node{$\mu$} (zero-dm1)
    (km1-dm1) edge[bend right, auto=left, dotted] node{$\lambda$} (k-dm1)
    (k-dm1) edge[bend right, auto=right, dotted] node{$K \mu$} (km1-dm1)
    (k-dm1) edge[bend left, auto=left, dotted] node{$\lambda$} (k-d)

    (zero-d) edge[bend right, auto=left, dotted] node{$\lambda$} (one-d)
    (one-d) edge[bend right, auto=right, dotted] node{$\mu$} (zero-d)
    (km1-d) edge[bend right, auto=left, dotted] node{$\lambda$} (k-d)
    (k-d) edge[bend right, auto=right, dotted] node{$K \mu$} (km1-d);
\end{tikzpicture}
\caption{System evolution within each deferral stage}
\label{fig:msmdfin}
\end{figure}

Finally, we are interested in characterizing the blocking probability
$P_{B},$ which is the long run fraction of customers dropped. This can
be done as follows:
\begin{equation}
  \label{eq:singledef_PB_1}
	P_{B} = \Pi_{1,1} + \frac{(\Pi_{0,1} +
			\Pi_{1,1})p_{1,1}(x)}{\lambda x}
\end{equation}
The first term above captures the long run fraction of customers
dropped on arrival (because the server was busy and there was already
a deferred customer). The second term captures the long run fraction
of customers dropped post deferral on rearrival. Indeed, the
probability that any deferred customer gets dropped on rearrival
equals $p_{1,1}(x).$ Thus, the rate of drops of deferred customers
equals $\frac{(\Pi_{0,1} + \Pi_{1,1})p_{1,1}(x)}{x}.$

To summarize, solving equations
\eqref{eq:ss-flow-1}--\eqref{eq:ss-tpr}, and applying the solution to
\eqref{eq:singledef_PB_1} yields the following result.
\begin{theorem}
  \label{thm:singledef}
  For the case $K = D = 1,$ under the DSRT policy, the blocking
  probability $P_B$ is given by 
  \begin{equation}
    \label{eq:singledef_PB}
    \text{P}_{B} = \frac{\rho}{1+\rho} - \frac{\rho(1-\text{exp}\{-(\lambda+\mu)x\})}{(1+\rho)^{2}(1+\rho(1+\lambda x))},
  \end{equation}
  where $\rho := \lambda/\mu$. 
\end{theorem}
Note that $\frac{\rho}{1+\rho}$ is the blocking probability of an
M/M/1/1 system. It is therefore clear from \eqref{eq:singledef_PB}
that the blocking probability under the DSRT policy is a strict
improvement over the scenario where deferral is not allowed. Moreover,
the improvement diminishes to zero when $x = 0$ and as $x \ra \infty.$
While the optimal value of $x$ that minimizes $P_B$ does not admit an
explicit formula, it can be shown that the optimal value is bounded
between $\frac{1}{\lambda+\mu}$ and $\frac{\sqrt{2}}{\lambda}$.

\subsection{Multiple-Servers, Multiple-Deferrals System}
\label{sec:mutiservermultideferral}

We now turn to the performance evaluation in the general setting,
i.e., the number of servers $K$ and the maximum number of deferrals
$D$ are arbitrary. Figure~\ref{fig:dyanmics_fig_2} illustrates the
temporal evolution of $\theta$ in this case. Note that each deferral
phase (an interval over which $d > 0$) is no longer necessarily of
duration $x$ anymore, but is instead made up of a random number of
\emph{deferral stages}, each of duration $x.$ Moreover, at the end of
each deferral phase, the configuration is of the form $(k,0)$ where
$0 < k \leq K.$ Thus, disregarding the deferral phases, the system
configuration evolves as per the Markov chain depicted in
Figure~\ref{fig:msmd_ctmc}. Here, $\alpha_i$ denotes the probability
that a deferral phase ends with $i$ active servers. (The values of
$\alpha_i$ are in turn dictated by the behavior of the system during
each deferral phase; we will characterize $\alpha_i$ shortly.)
Moreover, over each deferral stage of duration $x,$ the system
configuration evolves as per the Markov chain shown in
Figure~\ref{fig:msmdfin}.

In order to characterize $\alpha_i,$ and to arrive at the system of
equations relating the long run fractions of time spent in the
different configurations, we first analyse the evolution of the system
configuration over a single deferral stage, and then consider the
sequence of starting configurations across deferral stages within a
deferral phase.

\subsubsection{Evolution of system configuration within a single
  deferral stage}
\label{subsec:intrastage}

Deferral stages begin in a configuration of the form $(k,d),$ where
$1 \leq k \leq K,$ and $1 \leq d < D.$\footnote{Specifically, the
  first deferral stage in any deferral phase begins in configuration
  $(K,1)$.} Subsequently for duration $x,$ the system configuration
evolves as per the Markov chain depicted in
Figure~\ref{fig:msmdfin}. The transition rate matrix corresponding to
this chain is given by
%
%
\begin{equation*}
  Q = 
  \begin{bmatrix}
    Q_{1} & \Lambda & \textbf{0} & \textbf{0} & \ldots & \textbf{0} \\
    \textbf{0} & Q_{1} & \Lambda & \textbf{0} & \ldots & \textbf{0} \\    
    \textbf{0} & \textbf{0} & \ddots & \ddots  &  & \textbf{0} \\
    \textbf{0}& \textbf{0} & & \ddots & \ddots  & \textbf{0}\\
    \textbf{0} & \textbf{0} & \ldots & \textbf{0} & Q_{1} & \Lambda \\
    \textbf{0} & \textbf{0} & \textbf{0} & \ldots & \textbf{0} & Q_{2}
    \end{bmatrix},
\end{equation*}
where $\textbf{0}$ is an all-zero square matrix of size $(K+1)$,
$Q_{2}$ is the transition rate matrix of an M/M/K/K queue, $\Lambda$ is
a diagonal matrix with entries $(0,\cdots,0,\lambda)$, and
$Q_{1} = Q_{2} - \Lambda$. In writing this rate matrix, the
configurations $(k,d)$ have been ordered by stacking them row-wise
from Figure~\ref{fig:msmdfin}.

Considering the evolution of this chain over the interval $[0,x],$ the
distribution at time $t$ is denoted by the vector $\textbf{p}(t),$
where
%
\begin{equation*}
  \textbf{p}(t) = \textbf{p}_{\text{init}} e^{Qt}.
\end{equation*}
Here, the initial probability vector $\textbf{p}_{\text{init}}$ would
have an entry~1 according to the starting configuration of the stage
under consideration, the other entries being zero.

\subsubsection{Evolution of deferral stage starting configurations}
\label{sec:defstagestarts}

As mentioned before, the starting configuration in a deferral stage is
of the form $(k,d),$ where $1 \leq k \leq K,$ and $1 \leq d < D.$ We
construct a discrete time Markov chain to capture the evolution of the
starting configurations across successive deferral stages. To make the
chain recurrent, we add a dummy state $\mathcal{D}$ which captures the
end of the deferral phase.
Note that the transition from the dummy state is to the configuration
$(K,1),$ since that is the starting configuration on the next deferral
phase.

To describe the transition probability matrix corresponding to this
discrete time Markov chain, we need the following notation. Within a
deferral stage, let $p_{k',d'}^{k,d}(t)$ denote the probability of
being in configuration $(k', d')$ at time $t \in [0,x],$ with starting
configuration $(k,d).$ Note that these probabilities are defined in
Section~\ref{subsec:intrastage}.

Now, let $\hat{p}(c' | c)$ denote the probability that the starting
configuration in the next deferral stage is $c'$ given that the
starting configuration in the present deferral stage is $c,$ where
$$c,c' \in \hat{S} := \{(k,d)\: 1 \leq k \leq K,\ 1 \leq d < D \} \cup \{\mathcal{D}\}.$$ 
The transition probabilities $\hat{p}(\cdot | \cdot)$ are given by
\begin{align*}
  \hat{p}((k',d')|(k,d)) &= p_{k'-1, d'+1}^{k,d}(x) \\
  &\qquad\qquad \text{for } 1 \leq k' < K,\  1 \leq d' < D, \\
  \hat{p}((K,d')|(k,d)) &= p_{K-1, d'+1}^{k,d}(x) + p_{K, d'+1}^{k,d}(x) \\
  &\qquad\qquad \text{for } 1 \leq d' < D, \\
  \hat{p}(\mathcal{D}|(k,d)) &= \sum_{\ell = 0}^K p_{\ell,1}^{k,d}(x), \\
  \hat{p}((K,1)|\mathcal{D}) &= 1. \\
\end{align*}
Given these transition probabilities corresponding to this finite
irreducible discrete time Markov chain, one can obtain the stationary
distribution. Let $P_{\text{start}}^c$ denote the stationary
probability associated with configuration $c \in \hat{S}.$

\subsubsection{The equations relating $\Pi_{k,d}$}

We begin by relating the long run fractions of time spent in the
non-deferral configurations. This is in turn based on the Markov chain
depicted in Figure~\ref{fig:msmd_ctmc}. The probabilities $\alpha_l$ are
characterized as follows.
\begin{lemma}
  \label{lemma:alphai}
  For $1 \leq l < K,$
  $$\alpha_l = \frac{\sum_{j=1}^{k} P_{\text{start}}^{j,1}~
    p_{l-1,1}^{j,1}(x)}{\sum_{j=1}^{k} P_{\text{start}}^{j,1}~
    \sum_{l'=0}^{k}p_{l',1}^{j,1}(x)}.$$
\end{lemma}
\begin{IEEEproof}
  The characterization of $\alpha_l$ follows directly from the
  discrete-time Markov chain discussed in
  Section~\ref{sec:defstagestarts}; indeed, $\alpha_l$ is simply the
  probability that a deferral phase ends with a transition to the
  configuration $(l, 0).$ Consider a (discrete-time) renewal process
  where renewal instance correspond to visits to the dummy state
  (i.e., the end of a deferral phase) in the above DTMC. Each renewal
  cycle produces a reward of~1 if it ends with a transition to the
  configuration $(l, 0),$ and zero otherwise. An application of the
  renewal reward theorem yields
  $$\sum_{j=1}^{k} P_{\text{start}}^{j,1}~ p_{l-1,1}^{j,1}(x)
  = \alpha_l P_{\text{start}}^{\mathcal{D}}.$$ The statement
  of the lemma now follows, once we note that the balance equations
  for the DTMC under consideration imply
$$P_{\text{start}}^{\mathcal{D}} = \sum_{j=1}^{k}
P_{\text{start}}^{j,1}~ \sum_{l'=0}^{k}p_{l',1}^{j,1}(x).$$
\end{IEEEproof}
Given this result, we have:
\begin{equation}
  \label{eq:general_case1}
  \begin{array}{rl}
  \lambda \Pi_{0,0} &= \mu \Pi_{1,0}, \\
  (\lambda + l\mu) \Pi_{l,0} &= \lambda \Pi_{l-1,0} + (l+1)\mu\Pi_{l+1,0}
                               + \lambda \alpha_l \Pi_{K,0}, \\
 & \qquad \qquad \text{ for } (1 \leq l < K).
  \end{array}
\end{equation}

Next, we focus on the ratio of long run fraction of time spent in
deferral configurations. This is given by 
\begin{align}
  \label{eq:general_case2}
  \frac{\Pi_{k_{1},d_{1}}}{\Pi_{k_{2},d_{2}}} &= 
  \frac{\sum_{k=1}^{K} \sum_{d=1}^{D-1} 
  P_{\text{start}}^{k,d} \int_{0}^{x} p_{k_{1},d_{1}}^{k,d}(t)\text{d}t}
  {\sum_{k=1}^{K} \sum_{d=1}^{D-1} 
  P_{\text{start}}^{k,d} \int_{0}^{x} p_{k_{2},d_{2}}^{k,d}(t)\text{d}t},
\end{align}
where $0 \leq k_{1}, k_{2} \leq K$ and $1 \leq d_{1}, d_{2} \leq D$.

Finally, it remains to relate the time fractions corresponding to
deferral configurations to those corresponding to non-deferral
configurations. This is done as follows:
\begin{align}
  \label{eq:general_case3}
  \lambda \Pi_{K,0} = \frac{(\sum_{k=0}^{K}\sum_{d=1}^{D} \Pi_{k,d})
  (~\sum_{j=1}^{K} P_{\text{start}}^{j,1}~ \sum_{l=0}^{K}p_{l,1}^{j,1}(x)~)}{x(1 - P_{\text{start}}^{\mathcal{D}})}.
\end{align}
This is justified as follows.
$\lambda \Pi_{K,0}$ gives the rate of transition from non-deferral
configurations to deferral configurations. On the other hand, the rate
of rearrivals equals
$\frac{\sum_{k=0}^{K}\sum_{d=1}^{D} \Pi_{k,d}}{x}.$
$\frac{\sum_{j=1}^{K} P_{\text{start}}^{j,1}~
\sum_{l=0}^{K}p_{l,1}^{j,1}(x)}{1 - P_{\text{start}}^{\mathcal{D}}}$ 
gives the probability of exiting the
positive deferral configurations at the end of any deferral stage of
duration $x$ . \eqref{eq:general_case3} follows by combining these
arguments. 

Solving the system of linear
equations~\eqref{eq:general_case1}--\eqref{eq:general_case3} along
with the normalization condition yields $\Pi_{k,d}$ for
$(k,d) \in [K]\times[D].$

\subsubsection{Blocking probability characterization}

The blocking probability can be characterized in a similar manner as
before.
\begin{lemma}
  Under the DSRT policy,
\begin{align}
  P_{B} = \Pi_{K,D}
  + \Pi_{K,0}\frac{\sum_{k=1}^{K} \sum_{d=1}^{D-1}P^{k,d}_{\text{start}} 
  \sum_{l=1}^{D}p_{K,l}^{i,j}(x)}
  {\sum_{j=1}^{K} P_{\text{start}}^{j,1}~ \sum_{l=0}^{K}p_{l,1}^{j,1}(x)}.
  \label{eq:pb}
\end{align}  
\end{lemma}
\begin{IEEEproof}
  Customers are blocked in two ways. First, a customer arrives when
  the system configuration is $(K,D).$ Using the PASTA property, the
  long run fraction of arrivals blocked this way equals $\Pi_{K,D},$
  which accounts for the first term in \eqref{eq:pb}.

  Second, deferred customers can get blocked when all servers are busy
  upon rearrival. Let $DP(t)$ denote the number of deferral phases
  completed until time $t,$ and $B_i$ denote the number of deferred
  customers blocked in deferral phase~$i.$ Let $A(t)$ denote the
  number of arrivals until time~$t.$ The long run fraction of arrivals
  blocked upon rearrival equals
  \begin{align*}
    &\lim_{t \ra \infty} \frac{\sum_{i = 1}^{DP(t)} B_i}{A(t)}\\
    =& \lim_{t \ra \infty} \frac{t}{A(t)} \frac{DP(t)}{t} \frac{\sum_{i = 1}^{DP(t)} B_i}{DP(t)}\\
    =& \frac{1}{\lambda} \left( \lambda \Pi_{K,0} \right) \Exp{B_1}.  
  \end{align*}
  Finally, a renewal reward argument analogous to that in the proof of
  Lemma~\ref{lemma:alphai} yields
  \begin{align*}
    \Exp{B_1} &= \frac{\sum_{k=1}^{K} \sum_{d=1}^{D-1}P^{k,d}_{\text{start}}
                \sum_{l=1}^{D}p_{K,l}^{i,j}(x)}{P_{\text{start}}^{\mathcal{D}}} \\
    =&\frac{\sum_{k=1}^{K} \sum_{d=1}^{D-1}P^{k,d}_{\text{start}} 
  \sum_{l=1}^{D}p_{K,l}^{i,j}(x)}
  {\sum_{j=1}^{K} P_{\text{start}}^{j,1}~ \sum_{l=0}^{K}p_{l,1}^{j,1}(x)}.
  \end{align*}
\end{IEEEproof}


\subsection{Case Study}
\label{subsec:case_study_1}
In this section, we numerically compare the performance of the DSRT
policy with state-independent policies and the M/M/$K$/$K$ system in
the Halfin-Whitt regime (a.k.a. \emph{quality and efficiency driven}
regime; see~\cite{Halfin81}). We consider two simple state independent
policies: the first policy assigns a deferral time equal to $\hat{T}$
and the second policy assigns a deferral time uniformly sampled in the
interval $[0,\hat{T}]$. The Halfin-Whitt regime corresponds to a
sequence of queueing systems parameterized by the number of
servers~$K.$ The arrival rate in the system with $K$ servers equals
$K+\beta\sqrt{K}$, where $\beta$ is a constant. In our experiments,
the service rate for each server is set to be 1, the patience time
$\hat{T}$ is set to be 10, and $\beta$ is set to be 0.1.

\begin{figure}[htp]
    \centering
    \begin{subfigure}[b]{0.45\linewidth}
        \includegraphics[width=\linewidth]{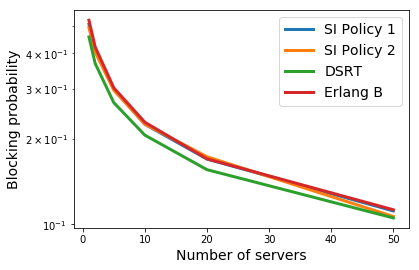}
        \caption{Maximum Deferrals: 1}
        \label{fig:pe1}
    \end{subfigure}
    \begin{subfigure}[b]{0.45\linewidth}
        \includegraphics[width=\linewidth]{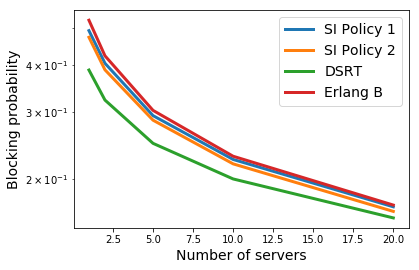}
        \caption{Maximum Deferrals: 2}
        \label{fig:pe2}
    \end{subfigure}
    \label{fig:pe_1_2}
    \caption{Performance Evaluation}
\end{figure}

We compare the policies for the cases when $\hat{D}$ equals~1 and~2.
We emphasize that $\hat{D}$ acts as a parameter for the
state-independent policies while the deferral limit for DSRT
$D = \min(\hat{D}, \floor{T/x})$, where $x \leq T$. For DSRT, the
blocking probability is evaluated numerically using the procedure
developed in previous sections and is optimized by tuning the
parameter~$x.$ For the state independent policies, we run MCMC
simulations where the blocking probability is calculated for 100,000
customers. The blocking probability for M/M/$K$/$K$ system is
calculated numerically using the Erlang-B formula. The blocking
probabilities of the policies for the cases where $\hat{D}$ is equal
to~1 and~2 are given in Figure~\ref{fig:pe1} and Figure~\ref{fig:pe2}
respectively. We note that DSRT outperforms the state-independent
policies, which produce a blocking probability close to the Erlang-B
formula.

\ignore{
Observe that the state dependent policy performs better than the state
independent policies. However, calculating and optimizing the blocking
probability for the state dependent policy is computationally expensive.
We now consider the cases when the number of deferrals is 5 and 10. 
Here, we perform MCMC simulations to find the optimal blocking 
probability for the state dependent policy as well. The evaluations are 
plotted in Figure~\ref{fig:pe5} and Figure~\ref{fig:pe10}. We can observe
that the state dependent policy performs better than the state independent 
policies for these cases as well. 

\begin{figure}[htp]
    \centering
    \begin{subfigure}[b]{0.45\linewidth}
        \includegraphics[width=\linewidth]{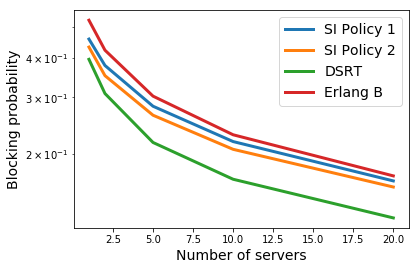}
        \caption{Maximum Deferrals: 5}
        \label{fig:pe5}
    \end{subfigure}
    \begin{subfigure}[b]{0.45\linewidth}
        \includegraphics[width=\linewidth]{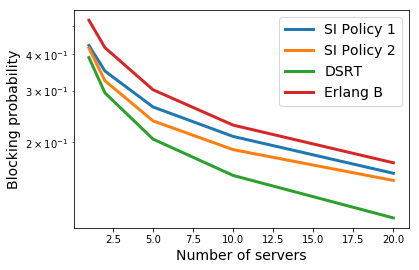}
        \caption{Maximum Deferrals: 10}
        \label{fig:pe10}
    \end{subfigure}
    \label{fig:pe_5_10}
    \caption{Performance Evaluation}
\end{figure}  }

\begin{figure}[htp]
    \centering
    \includegraphics[width=0.7\linewidth]{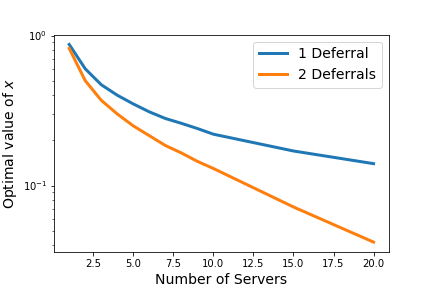}
    \caption{Optimal $x$ for DSRT}
    \label{fig:opt_x_sd}
\end{figure}

The optimal value of $x$ is plotted against the number of servers in
Figure~\ref{fig:opt_x_sd}. Note that as the number of servers
increases, the optimal value of $x$ shrinks. Characterizing this
behavior analytically presents an interesting avenue for future work.

\section{Deferral Times Bounded in Expectation}
\label{sec:model2}

In this section, we consider a relaxation of the deferral constraint
considered before, in that the upper bound on deferral times is
imposed \emph{in expectation} (rather than on each instance, as in
Section~\ref{sec:sd_policy}). Specifically, we require that the
expected deferral time should not exceed $\hat{T}$. This relaxation 
may be interpreted as an approximation of the `hard' deferral constraint
considered before.

In the above setting, we propose and analyse a deferral policy,
referred to as the {\bf E}xponentially {\bf S}paced {\bf R}earrival
{\bf T}imes (ESRT) policy. As the name suggests, this policy
introduces an exponentially distributed spacing between rearrival
instants. Specifically, the ESRT policy allots exponentially separated
deferral times, i.e., the difference between the rearrival times of
deferred arrivals is exponential with rate $\alpha$, where
$1/\alpha \leq \hat{T}$, the deferral limit being
$D = \min(\hat{D}, \floor{\hat{T}\alpha});$ this policy is described
formally in Algorithm~\ref{algesrt}. Note that $\text{Exp}(\alpha)$
refers to an exponential random variable with rate $\alpha$ that is
independent of the workload, as well as other inter-rearrival
intervals.
\begin{algorithm}[htp]
  \caption{ESRT Policy}
  \label{algesrt}
\begin{algorithmic}
  \Procedure{}{}
  \If{$\text{num\_deferrals} = 0$}
    \State $\text{rearrival\_time} = \text{arrival\_time} + \text{Exp}(\alpha)$
  \ElsIf{$0 < \text{num\_deferrals} < D$}
    \State $\text{rearrival\_time} = \text{prev\_rearrival\_time} + \text{Exp}(\alpha)$
  \Else
    \State $\text{customer blocked}$
  \EndIf
  \EndProcedure
\end{algorithmic}
\end{algorithm}

\begin{figure}[htp]
\centering
\begin{tikzpicture}[state/.style={circle, draw, minimum size=1.4cm}]
\node[state] (zero-zero) {0,0};
\node[state, right=of zero-zero] (one-zero) {1,0};
\node[state, right=of one-zero] (km1-zero) {K-1,0};
\node[state, right=of km1-zero] (k-zero) {K,0};

\node[state, below=of zero-zero] (zero-one) {0,1};
\node[state, below=of one-zero] (one-one) {1,1};
\node[state, right=of one-one] (km1-one) {K-1,1};
\node[state, right=of km1-one] (k-one) {K,1};

\node[state, below=of zero-one] (zero-dm1) {0,D-1};
\node[state, below=of one-one] (one-dm1) {1,D-1};
\node[state, right=of one-dm1] (km1-dm1) {K-1,D-1};
\node[state, right=of km1-dm1] (k-dm1) {K,D-1};

\node[state, below=of zero-dm1] (zero-d) {0,D};
\node[state, below=of one-dm1] (one-d) {1,D};
\node[state, right=of one-d] (km1-d) {K-1,D};
\node[state, right=of km1-d] (k-d) {K,D};

\path (one-zero) -- node[auto=false]{\ldots} (km1-zero);
\path (one-one) -- node[auto=false]{\ldots} (km1-one);
\path (one-dm1) -- node[auto=false]{\ldots} (km1-dm1);
\path (one-d) -- node[auto=false]{\ldots} (km1-d);

\path (zero-one) -- node[auto=false]{\vdots} (zero-dm1);
\path (one-one) -- node[auto=false]{\vdots} (one-dm1);
\path (km1-one) -- node[auto=false]{\vdots} (km1-dm1);
\path (k-one) -- node[auto=false]{\vdots} (k-dm1);

\draw[every loop]
  (zero-zero) edge[bend right, auto=left] node{$\lambda$} (one-zero)
  (one-zero) edge[bend right, auto=right] node{$\mu$} (zero-zero)
  (km1-zero) edge[bend right, auto=left] node{$\lambda$} (k-zero)
  (k-zero) edge[bend right, auto=right] node{$K \mu$} (km1-zero)
  (k-zero) edge[bend left, auto=left] node{$\lambda$} (k-one)
  
  (zero-one) edge[auto=left] node{$\alpha$} (one-zero)
  (km1-one) edge[auto=left] node{$\alpha$} (k-zero)
  (k-one) edge[auto=left] node{$\alpha$} (k-zero)
  
  (zero-one) edge[bend right, auto=left] node{$\lambda$} (one-one)
  (one-one) edge[bend right, auto=right] node{$\mu$} (zero-one)
  (km1-one) edge[bend right, auto=left] node{$\lambda$} (k-one)
  (k-one) edge[bend right, auto=right] node{$K \mu$} (km1-one)
  
  (zero-dm1) edge[bend right, auto=left] node{$\lambda$} (one-dm1)
  (one-dm1) edge[bend right, auto=right] node{$\mu$} (zero-dm1)
  (km1-dm1) edge[bend right, auto=left] node{$\lambda$} (k-dm1)
  (k-dm1) edge[bend right, auto=right] node{$K \mu$} (km1-dm1)
  (k-dm1) edge[bend left, auto=left] node{$\lambda$} (k-d)
  
  (zero-d) edge[auto=left] node{$\alpha$} (one-dm1)
  (km1-d) edge[auto=left] node{$\alpha$} (k-dm1)
  (k-d) edge[auto=left] node{$\alpha$} (k-dm1)
  
  (zero-d) edge[bend right, auto=left] node{$\lambda$} (one-d)
  (one-d) edge[bend right, auto=right] node{$\mu$} (zero-d)
  (km1-d) edge[bend right, auto=left] node{$\lambda$} (k-d)
  (k-d) edge[bend right, auto=right] node{$K \mu$} (km1-d);
\end{tikzpicture}
\caption{Dynamics of ESRT}
\label{fig:esrt}
\end{figure}
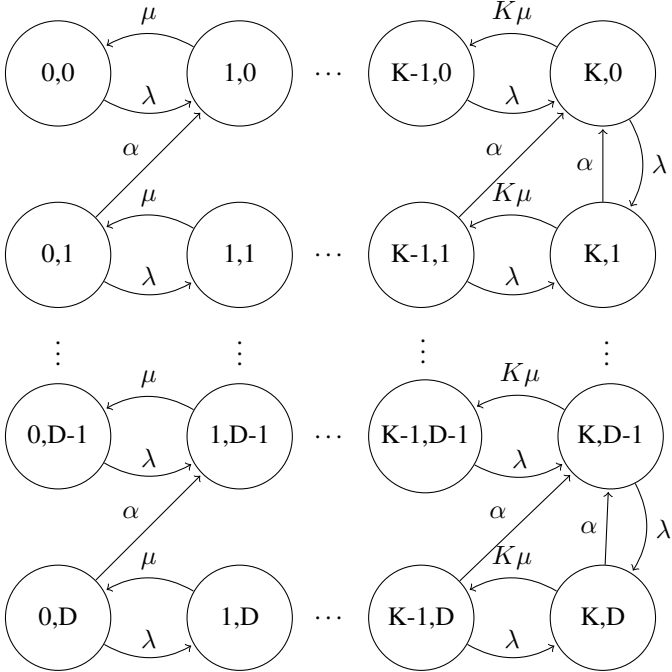

Given the memorylessness of the interarrival times, service times, and
the time until next rearrival, the evolution of the system under the
ESRT policy is described by a continuous time Markov chain over the
state space $[K]\times[D],$ where state $(k,d)$ indicates that $k$
servers are currently active, and there are $d$ jobs in deferral. The
transition rate diagram for this Markov chain is depicted in
Figure~\ref{fig:esrt}. Note that memorylessness of the time until
the next rearrival results in a much simpler system description
(compared to the DSRT policy analysed in the previous section), and
yields a much more explicit performance evaluation. In the following,
we first compute the stationary distribution corresponding to the ESRT
Markov chain, and then use that to characterize the blocking
probability.

\subsection{Obtaining the stationary distribution}

Matrix geometric methods can be used to obtain the stationary
distribution corresponding to the Markov chain shown in
Figure~\ref{fig:esrt}. Let $\Pi_{k,d}$ represent the steady state
probability of state $(k,d).$

We begin by writing the flow balance equations for states with $D$
deferrals
\begin{align*}
  &(\lambda + \alpha) \Pi_{0,D} = \mu \Pi_{1,D}, \\
  &(\lambda + j\mu + \alpha) \Pi_{j,D} = 
  (j+1)\mu\Pi_{j+1,D} + \lambda \Pi_{j-1,D}, \\
  &\hspace{2.5in}\text{for } (0<j<K).
\end{align*}
Letting $\vec{\Pi}_{D} := [\Pi_{0,D},\cdots,\Pi_{K,D}]$, the above
system of $K$ equations can be used to characterize $\vec{\Pi}_{D}$
upto a multiplicative constant. Specifically, we write
\begin{equation}
  \label{eq:esrt1}
  \vec{\Pi}_{D} = c \vec{V}_{D}, 
\end{equation}
where $\vec{V}_{D}$ is a row vector whose last entry is 1.

Now, consider the collection of states corresponding to $d$ deferrals,
where $d \in \{1,\cdots,D-1\}$ deferrals. The flow balance equations
for these states are as follows
\begin{align*}
  &(\lambda + \alpha) \Pi_{0,d} = \mu \Pi_{1,d}, \\
  &(\lambda + j\mu + \alpha) \Pi_{j,d} = 
  (j+1)\mu\Pi_{j+1,d} + \lambda \Pi_{j-1,d} + \alpha \Pi_{j-1,d+1}, \\
  &\hspace{2.5in} \text{for } (0<j<K), \\
  & \lambda \Pi_{K, d} = \alpha \sum_{i=0}^{K} \Pi_{i,d+1}.
\end{align*}
Let $\vec{\Pi}_{d} = [\Pi_{0,d},\cdots,\Pi_{K,d}]$, then the above
equations can be written as
\begin{align*}
  &\vec{\Pi}_{d} A(\alpha) = \vec{\Pi}_{d+1}B(\alpha) \text{ where } \\
  &A(\alpha) = 
  \begin{bmatrix}
    \lambda + \alpha & -\lambda & 0 & \cdots & 0 \\
    -\mu & \lambda + \mu + \alpha & -\lambda & \cdots & 0 \\
    0 & -2\mu & \lambda + 2\mu + \alpha & \cdots & 0 \\
    \vdots & \vdots & \ddots & \vdots & \vdots \\
    0 & 0 & \cdots & -D\mu & \lambda  
  \end{bmatrix} \\
  \text{and }&B(\alpha) = 
  \begin{bmatrix}
    0 & \alpha & 0 & \cdots & \alpha \\
    0 & 0 & \alpha & \cdots & \alpha \\
    0 & 0 & 0 & \cdots & \alpha \\
    \vdots & \vdots & \ddots & \vdots & \vdots \\
    0 & 0 & 0 & \cdots & \alpha
  \end{bmatrix}.
\end{align*}  
One can show using a simple manipulation and strict diagonal dominance
that $A(\alpha)$ is non-singular.
Hence, we have the following recursive relation
\begin{align}
  \label{eq:esrt2}
  \vec{\Pi}_{d} = \vec{\Pi}_{d+1}R(\alpha) \quad \text{ for } 1 \leq d < D,  
\end{align}
where $R(\alpha)=B(\alpha) A(\alpha)^{-1}.$

Finally, considering the flow balance equations for states with no
deferrals,
\begin{align*}
  &\lambda \Pi_{0,0} = \mu \Pi_{1,0}, \\
  &(\lambda+j\mu) \Pi_{j,0} = \lambda \Pi_{j-1,0} + (j+1)\mu \Pi_{j+1, 0} 
  + \alpha \Pi_{j-1,1}, \\
  &\hspace{2.5in}\text{for } (0<j<K), \\
  & \lambda \Pi_{K, 0} = \alpha \sum_{i=0}^{K} \Pi_{i,1}.
\end{align*}
The equations above can be written as
$\vec{\Pi}_{0} A(0) = \vec{\Pi}_{1} B(\alpha),$ implying
\begin{align}
  \label{eq:esrt3}
  \vec{\Pi}_{0} &= \vec{\Pi}_{1} R_{1}(\alpha)
\end{align} 
where $R_{1}(\alpha) = B(\alpha) A(0)^{-1}$.\footnote{The
  invertibility of $A(0)$ follows since the determinant of $A(0)$ is
  $\lambda^{K+1}$.}

Using \eqref{eq:esrt1}--\eqref{eq:esrt3}, the stationary distribution
is completely determined upto the multiplicative constant $c,$ which
can be obtained by applying the law of total probability:
\begin{align*}
  c \vec{V}_{D} \left[R_{1}R^{D-1} + (I-R)^{-1}(I-R^{D})\right] \mathbbm{1} &= 1
\end{align*}

\subsection{The blocking probability}

Having characterized the stationary distribution associated with the
system state, we now express the blocking probability in terms of the
stationary distribution. As before, a job may be blocked either on
arrival into state $(K,D),$ or on rearrival if all servers are
occupied. Since $\alpha \sum_{d=1}^{D} \Pi_{K, d}$ is the rate of
dropped rearrivals, an elementary renewal argument shows that the long
run fraction of jobs dropped on rearrival equals
$\frac{\alpha}{\lambda} \sum_{d=1}^{D} \Pi_{K, d}.$ Thus,
\begin{align*}
  P_{B} = \Pi_{K,D} + \frac{\alpha}{\lambda} \sum_{d=1}^{D} \Pi_{K, d}. 
\end{align*}

For the special case $K = 1,$ the blocking probability takes the
following explicit form:
\begin{align*}
  P_{B}(1, D) = \frac{\lambda}{\lambda+\mu} \times
\frac{(ab)^{D+1} + (b-1)(ab)^{D} - b}{ab - 1 + a \frac{\lambda}{\lambda+\mu} [(ab)^D -1]}
\end{align*} 
where $a = \frac{\lambda}{\alpha}$ and
$b = \frac{\lambda+\alpha}{\lambda+\alpha+\mu}$.
\ignore{
If we further assume
$D=1,$ this simplifies to 
\begin{align*}
  P_{B}(1,1) = \frac{\lambda}{\lambda+\mu} \times \frac{1}{1 + \frac{\alpha \mu^2}{(\lambda+\mu)(\lambda+\alpha)^2} }.
\end{align*}
One can easily check that the optimal value of $\alpha$ in the case
above turns out to be $\lambda$. When there is exactly one server and
a maximum of one deferral, the optimal value of deferral rate doesn't
depend on the service rate and depends just on the arrival rate.}

\subsection{Case Study}
\label{subsec:case_study_2}

\begin{figure}[htp]
    \centering
    \begin{subfigure}[b]{0.45\linewidth}
        \includegraphics[width=\linewidth]{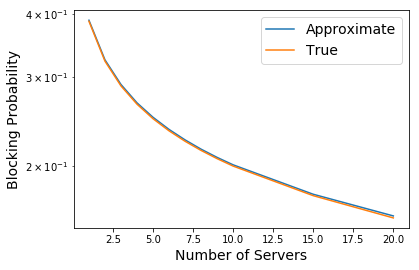}
        \caption{Approximate $P_{B}$}
        \label{fig:x_opt_approx_2}
    \end{subfigure}
    \begin{subfigure}[b]{0.45\linewidth}
        \includegraphics[width=\linewidth]{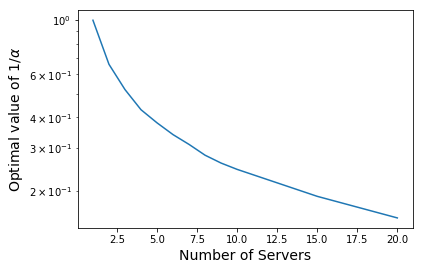}
        \caption{Optimal $\frac{1}{\alpha}$ for ESRT}
        \label{fig:x_opt_vals}
    \end{subfigure}
    \label{fig:x_opt_approx}
\end{figure}

Given that the ESRT policy is more tractable than the DSRT policy, our
first goal is to check if the performance of the former is a good
approximation to the performance of the latter. The numerical analysis
is done for the case when $\hat{D}=2$, and $\hat{T}=10$. i.e., the
system is allowed to defer one or two customers. We then consider a
sequence of queueing systems under the Halfin-Whitt scaling regime,
with $\beta = 0.1.$ In Figure~\ref{fig:x_opt_approx_2}, we compare the
blocking probability under ESRT using the optimal computed value
$\alpha^*(K)$ of the policy parameter, with the blocking probability
under DSRT using $x = 1/\alpha^*(K).$\footnote{Note that it is natural
  to approximate the performance under DSRT with parameter $x,$ with
  the performance under ESRT with $\alpha = 1/x.$} Note that the two
curves are nearly indistinguishable, suggesting that the approximation
is sound (at least for small values of $\hat{D}$). In
Figure~\ref{fig:x_opt_vals}, we plot $1/\alpha^*(K)$ v/s $K.$ Since
$1/\alpha^*(K)$ decreases with $K,$ it follows that it is optimal to
decrease the spacing between rearrivals as the size of the system
grows under the Halfin-Whitt regime. This is to be expected, since the
arrival rate and the service rate of the system grow nearly
proportionately under this scaling regime.

\begin{figure}[htp]
    \centering
    \begin{subfigure}[b]{0.45\linewidth}
        \includegraphics[width=\linewidth]{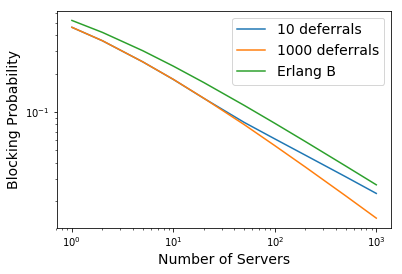}
        \caption{Performance Evaluation}
        \label{fig:esrt_erb}
    \end{subfigure}
    \begin{subfigure}[b]{0.45\linewidth}
        \includegraphics[width=\linewidth]{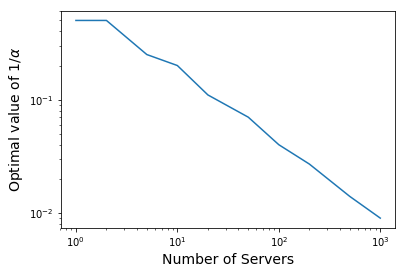}
        \caption{Optimal $\frac{1}{\alpha}$ for ESRT}
        \label{fig:x_opt_vals_1000}
    \end{subfigure}
    \label{fig:large_sys_anal}
\end{figure}

Next we study the impact of the maximum number of deferrals $\hat{D}.$ Under
the Halfin-Whitt scaling regime as before, in
Figure~\ref{fig:esrt_erb}, we plot the optimal (with respect to the
deferral parameter $\alpha$) blocking probability when $\hat{D}=10$ and
$\hat{D}=1000,$ along with the Erlang B formula (which corresponds to
$\hat{D} = 0.$) For this case, we set $\hat{T}=1.$ 
Note that increasing the limit on the number of deferrals
does improve the performance significantly. Moreover, it is well-known
that the blocking probability of the Erlang B (M/M/$K$/$K$) system
decays as $\Omega(1/\sqrt{K})$ under the Halfin-Whitt scaling
(see \cite{jagerman1974}). The numerical results as shown in
Figure~\ref{fig:esrt_erb}, suggest that the decay of blocking
probability of ESRT is also $\Omega(1/\sqrt{K})$ (note the nearly
linear scaling of blocking probability with $K$ on a log-log
scale), at least when $\hat{D}$ is small. Having a large value of
$\hat{D}$ could possibly allow a $o(1/\sqrt{K})$ decay in the
blocking probability. Formalizing this scaling behavior is an 
interesting avenue for future work.

When $\hat{D}$ is 1000, we also plot the inverse of optimal deferral
parameter $\alpha$ against the number of servers in
Figure~\ref{fig:x_opt_vals_1000}. Notice the nearly linear decay of
$1/\alpha$ in the log-log plot, indicating a regularly varying scaling
with the number of servers in the Halfin-Whitt regime.  Characterizing
the optimal deferral parameters for both ESRT and DSRT would also be
illuminating.

\section{Concluding Remarks}
\label{sec:conclusion}
Our work motivates several directions for future work including the
following. One, characterizing the optimal parameter values and
optimal blocking probabilities in a large system limit for DSRT/ESRT.
Second, designing deferral policies operating under constraints on the
number of deferrals and patience time that can beat the $1 / \sqrt{K}$
blocking probability scaling under the Halfin-Whitt regime. Finding
fundamental lower bounds on the blocking probability of such deferral
policies would also be interesting.


\small{
\bibliography{references}
}

\end{document}